\documentclass[aps,prl,reprint,superscriptaddress,amsmath,amssymb,nofootinbib,longbibliography]{revtex4-2}

\usepackage{graphicx}
\usepackage{dcolumn}
\usepackage{bm}
\usepackage{xcolor}
\usepackage{xspace}
\usepackage[colorlinks=true,linkcolor=blue,citecolor=blue,urlcolor=blue]{hyperref}
\usepackage[capitalise]{cleveref}

\newcommand{\gev}{\ensuremath{\mathrm{GeV}}}

\newcommand{\tev}{\ensuremath{\mathrm{TeV}}}

\newcommand{\Umu}{\ensuremath{|U_{\mu N}|^2}\xspace}
\newcommand{\Ue}{\ensuremath{|U_{e N}|^2}\xspace}
\newcommand{\Ual}{\ensuremath{\mathcal{U}_\alpha^2}\xspace}   % pair-summed flavor mixing
\newcommand{\abinv}{\ensuremath{\,\mathrm{ab}^{-1}}}
\begin{document}

\title{Probing the Origin of Neutrino Mass with Intense Muon Beams}
%\title{Probing Neutrino Mass and the Matter–Antimatter Puzzle with Intense Muon Beams}
%\title{Muon Beams at the Seesaw Frontier}
%\title{Muon Beams to Probe Neutrino Mass Generation}
%\title{Novel Heavy Neutral Lepton Searches With Next-Generation Muon Accelerators}

\author{Carlos A. Arg\"uelles}
%\email{carguelles@g.harvard.edu}
\affiliation{Department of Physics \& Laboratory for Particle Physics and Cosmology, Harvard University, Cambridge, MA 02138, USA\looseness=-1}
\author{Gonzalo Herrera}
%\email{gherreramoreno@fas.harvard.edu}
\affiliation{Kavli Institute for Astrophysics and Space Research, Massachusetts Institute of Technology, Cambridge, MA 02139, USA\looseness=-1}
\affiliation{Department of Physics \& Laboratory for Particle Physics and Cosmology, Harvard University, Cambridge, MA 02138, USA\looseness=-1}
\author{Nicholas Kamp}
\email{nkamp@fas.harvard.edu}
\affiliation{Department of Physics \& Laboratory for Particle Physics and Cosmology, Harvard University, Cambridge, MA 02138, USA\looseness=-1}
\author{Peng-Cheng Lu}
\email{pclu@sdu.edu.cn}
\affiliation{School of Physics, Shandong University, Jinan, Shandong 250100, China\looseness=-1}
\affiliation{Department of Physics and Astronomy, Michigan State University, East Lansing, MI 48824, USA}
\author{Keping Xie}
\email{kepingxie@sjtu.edu.cn}
\affiliation{State Key Laboratory of Dark Matter Physics \& Key Laboratory for Particle Astrophysics and Cosmology (MOE) \& Shanghai Key Laboratory for Particle Physics and Cosmology, Tsung-Dao Lee Institute \& School of Physics and Astronomy, Shanghai Jiao Tong University, Shanghai 201210, China\looseness=-1}
\affiliation{Department of Physics and Astronomy, Michigan State University, East Lansing, MI 48824, USA}

\begin{abstract}
Heavy neutral leptons (HNLs) can explain the origin and smallness of neutrino masses through a low-scale seesaw, but the smallest allowed mixings remain beyond existing experiments.
To address this, we develop three HNL search strategies using intense proposed multi-\tev\ muon beams: a heavy neutrino factory, a muon beam dump, and a muon--ion collider.
Together, their sensitivities reach at least two orders of magnitude beyond existing limits for HNL masses throughout the majority of the $1~{\rm MeV} \lesssim m_N \lesssim 100~{\rm GeV}$ range.
%,spanning MeV masses to the electroweak scale.
%Dense or extended targets enhance HNL production, while evacuated decay volumes suppress backgrounds from neutrino interactions.
Our muon beam dump strategy in particular reaches up to six orders of magnitude over existing limits at $m_N \sim 10~{\rm GeV}$, extending beyond the projected sensitivity of the $e^+e^-$ mode of the Future Circular Collider by over an order of magnitude.
We propose three beam dump configurations that are sensitive to mixings predicted by the type~I seesaw model for $10\lesssim m_N/{\rm GeV} \lesssim 50$, the first existing or proposed experiment to reach this target for HNL masses above the kaon mass.
%Our muon-beam-based HNL searches are the first to be sensitive to mixings predicted by the type~I seesaw model for $m_N > 1~{\rm GeV}$.
Moreover, they cover a significant fraction of the parameter space for which HNLs can explain the matter-antimatter asymmetry of the Universe. %Leptogenesis in the early Universe.
%Thus, modest extensions to a future muon accelerator facility can shed light on the origin of neutrino masses and the origin of matter in the Universe.
\end{abstract}

\maketitle

\emph{Introduction.---}%
The origin of neutrino mass and the Universe's excess of matter over antimatter remain unexplained.
Heavy neutral leptons (HNLs), gauge-singlet fermions that mix with the active neutrinos, can address both questions~\cite{Abdullahi:2022jlv}. 
They generate light-neutrino masses through the type-I seesaw~\cite{Minkowski:1977sc,Mohapatra:1979ia,Gell-Mann:1979vob,Yanagida:1979as}.
CP-violating oscillations of a nearly degenerate HNL pair can produce unequal abundances of leptons and antileptons in different flavors in the early Universe~\cite{Akhmedov:1998qx,Asaka:2005pn}. 
Thermal electroweak processes known as sphaleron transitions partially convert this asymmetry into an excess of baryons over antibaryons~\cite{Kuzmin:1985mm,Fukugita:1986hr}.

Low-scale realizations of type-I seesaw bring the HNL masses within experimental reach~\cite{Gorbunov:2007ak,Atre:2009rg,Ibarra:2011xn,Drewes:2016jae}, but testing the smallest allowed mixings remains difficult~\cite{Drewes:2019mhg}.
In the minimal seesaw with two quasi-degenerate HNLs of mean mass $m_N$, the total mixing of the heavy states with all active flavors obeys $U^2\geq\sum_i m_i/m_N\simeq6\times10^{-11}\,(\gev/m_N)$ for the normal ordering~\cite{Drewes:2016jae}.
This bound defines the total seesaw floor; the per-flavor mixing floors derived below can be substantially lower.
Above a GeV these targets lie orders of magnitude beneath the reach of existing fixed-target, collider, and meson-decay searches.
Small mixings suppress HNL production and lengthen their lifetimes.
Testing this region therefore requires an intense source and a decay region suited to long-lived particles.

Next-generation muon accelerators can address this challenge.
The beam energy envisaged for next-generation muon facilities opens production of heavier HNLs, while the expected intensity opens access to rare interactions. 
The proposed muon-collider stages deliver $E_\mu=1.5$ and $5~\tev$ beams~\cite{Accettura:2023ked,InternationalMuonCollider:2025sys,AlAli:2021let,InternationalMuonCollider:2024jyv,Begel:2025ldu} with an integrated intensity of order $10^{22}$ muons~\cite{Denton:2024glz}.
Charged-current scattering of such a beam on a fixed target produces HNLs up to $m_N\sim\sqrt{2m_pE_\mu}\approx100~\gev$ at $5~\tev$, in a collimated, highly boosted flux.
The decays of the same muons also enable a broad neutrino-factory program~\cite{Geer:1997iz,Bojorquez-Lopez:2024bsr,Mangano:2001mj,Denton:2024glz,Choi:2026yzw,Burk:2026fox}; here we exploit the muon beam itself as an HNL source.

\begin{figure*}[t]
  \centering \includegraphics[width=1.05\textwidth]{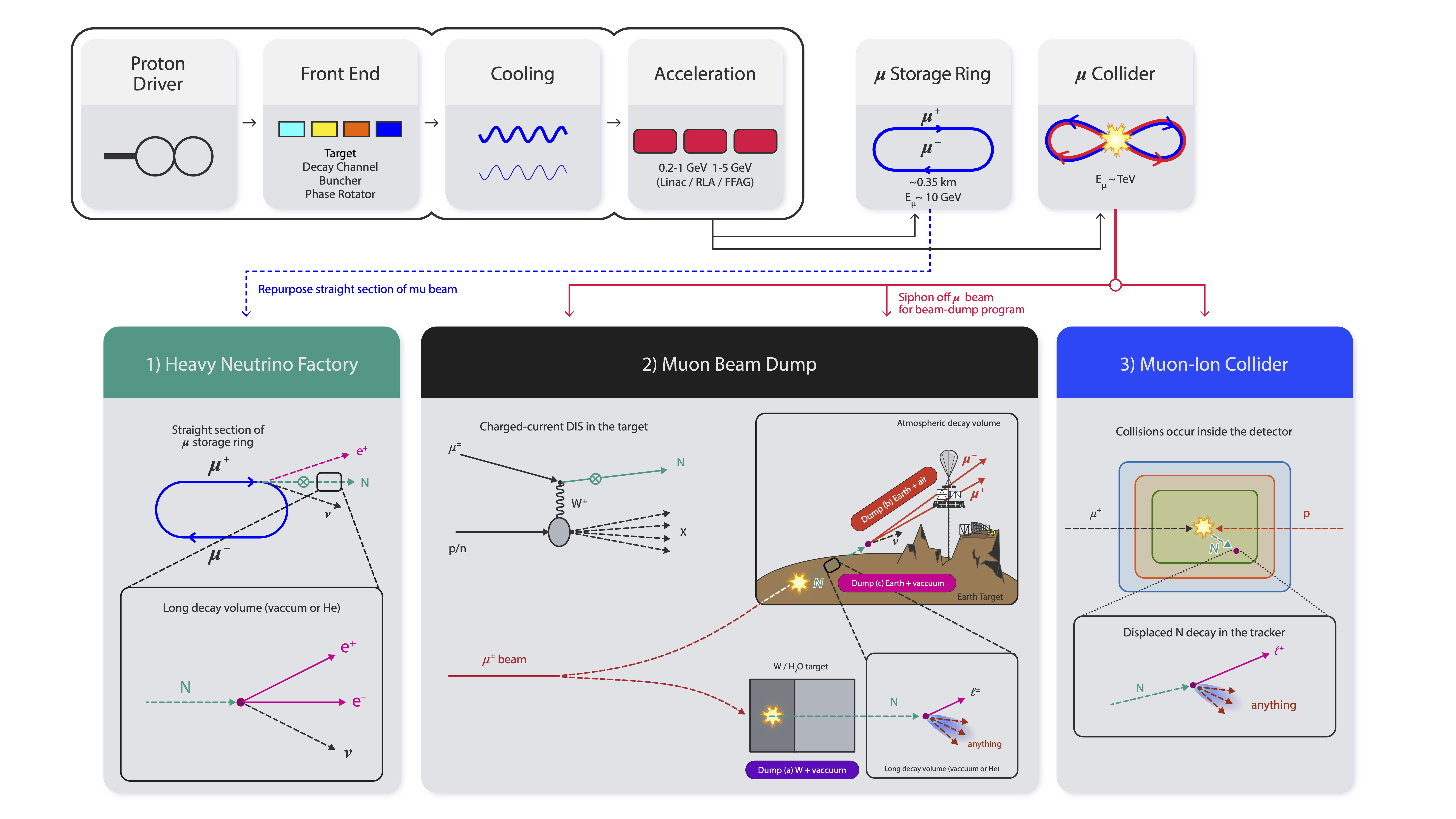}
  \caption{The three HNL search proposals enabled by an intense muon beam: a heavy neutrino factory using storage-ring muon decays, a muon beam dump, and a muon--ion collider.
  %\kp{(1) Unify ``Muon-Proton" $\to$ "Muon-Ion"; (2) $\mu^\pm p/n\to N+X$; (3) Siphon \textbf{off}; (4) $E_\mu\sim 10~\textrm{GeV/TeV}$.}. 
  The beam dump is studied with an earth target and atmospheric detection, a dense target and vacuum decay volume, and an earth target and vacuum decay tunnel.}
  \label{fig:diagram}
\end{figure*}

\begin{figure*}[t]
  \centering
  \includegraphics[width=0.49\textwidth]{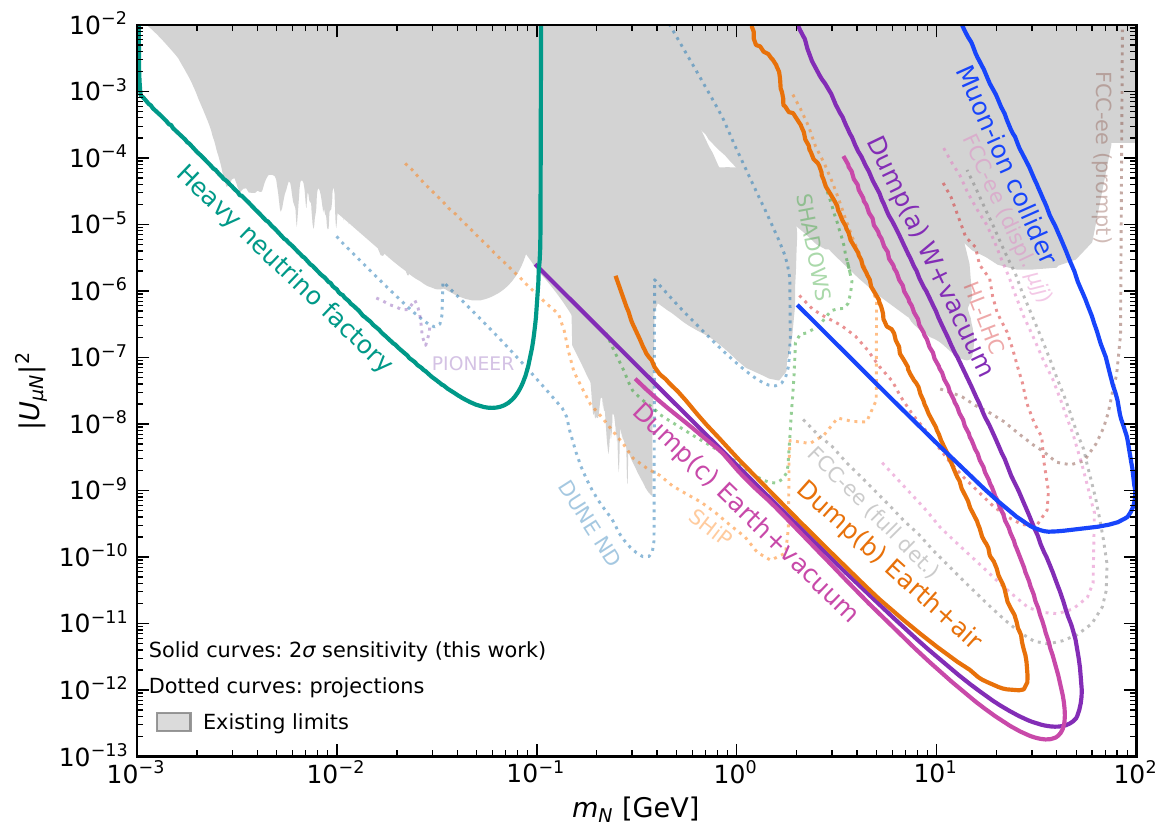}\hfill
  \includegraphics[width=0.49\textwidth]{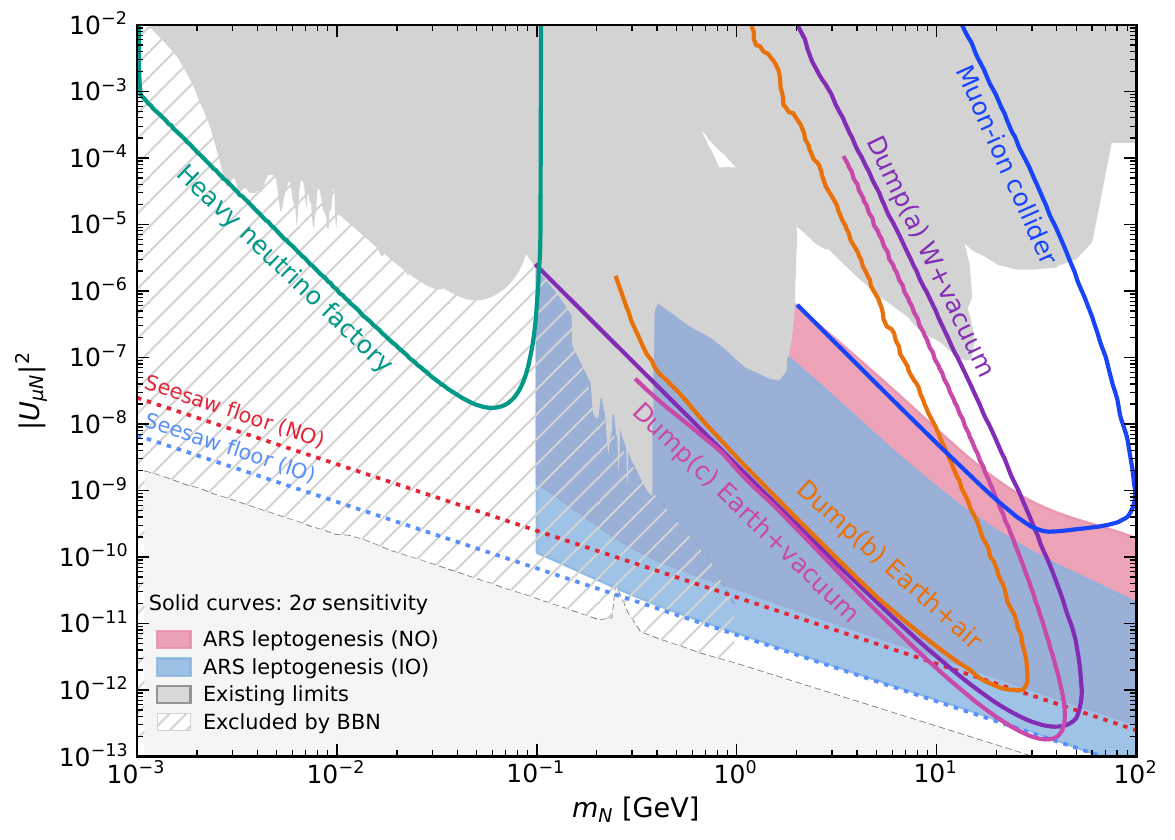}
  \caption{Projected $2\sigma$ sensitivities (solid) of the heavy neutrino factory (teal), the muon beam dump in its tungsten-plus-vacuum [Dump (a), purple], Earth-plus-atmosphere [Dump~(b), orange], and Earth-plus-vacuum [Dump~(c), magenta] configurations, and the muon--ion collider (blue).
  %Dump benchmarks use $E_\mu=5~\tev$ and $N_\mu=10^{22}$; the collider uses $10\abinv$.
  Gray regions are existing limits~\cite{Fernandez-Martinez:2023phj}. Left: comparison with projections (dotted) for DUNE~\cite{Breitbach:2021gvv}, SHADOWS~\cite{Baldini:2021shadows}, SHiP~\cite{SHiP:2018xqw}, PIONEER~\cite{PIONEER:2022yag}, HL-LHC~\cite{Drewes:2019fou}, and FCC-ee~\cite{Blondel:2022qqo,Bellagamba:2025xpd}. Right: comparison with the theoretical targets; hatching denotes the BBN constraint~\cite{Sabti:2020yrt}, dotted lines the muon-flavor seesaw floors of Eq.~\eqref{eq:flavor_floor}, and shaded bands the projected ARS leptogenesis targets for NO (red) and IO (blue) at the NuFIT~5.3 best fit.}
  \label{fig:combined}
\end{figure*}

We put forward three strategies, illustrated in Fig.~\ref{fig:diagram}: a storage-ring \emph{heavy neutrino factory}, a \emph{muon beam dump}, and a \emph{muon--ion collider}.
The beam dump is studied in three target--decay-region configurations, labeled as in Fig.~\ref{fig:combined}: (a) tungsten plus vacuum, (b) the Earth plus the atmosphere, and (c) the Earth plus vacuum.
In this Letter, we present the HNL sensitivity of each proposal, which in combination would improve upon existing limits by up to several orders of magnitude across $10^{-3} < m_N/{\rm GeV} < 10^2$, covering a substantial region of the parameter space for which HNLs can produce a primordial lepton asymmetry.
The muon beam dump proposals in particular would be the first experiments sensitive to the seesaw floor in \Umu.
\Cref{fig:combined} summarizes the reach of the three proposed HNL searches.
Further details on each are given in the companion article~\cite{Arguelles:2026companion}.

\emph{Production, decay, and theoretical targets.---}%
%The experimental projections assume a single HNL with pure muon mixing \Umu, except where electron mixing is discussed.
HNLs with nonzero muon mixing \Umu can be produced from muon decays and scattering.
Below $m_\mu$, HNLs can be produced in $\mu^-\to e^-\bar\nu_eN$~\cite{Ema:2023buz}.
At all masses up to the kinematic limit $m_N^{\rm max} \approx \sqrt{2 E_\mu m_{p/n} }$ they are produced in scattering off of nucleons, $\mu^\pm (p/n)\to NX$, through $t$-channel $W$ exchange.
We compute the deep-inelastic cross sections per nucleon with \texttt{MadGraph5\_aMC@NLO}~\cite{Alwall:2014hca} using CT18NNLO~\cite{Hou:2019efy} and an isoscalar target, and integrate them along the muon's path as it loses energy via ionization and radiative processes.
The cross sections are nearly mass-independent at low $m_N$ and fall toward the kinematic limit.
Decays proceed via $W$ and $Z$ bosons to leptonic and hadronic final states, with more available channels as the HNL mass increases~\cite{Arguelles:2026companion}.
% For a mean decay length $\ell_d=\beta\gamma c\tau_N$, the probability to decay between distances $L_0$ and $L_1$ along the trajectory is
% \begin{equation}\label{eq:decay_probability}
%     P_{\rm dec}=e^{-L_0/\ell_d}-e^{-L_1/\ell_d}.
% \end{equation}
Below the kinematic limit production scales as \Umu, and in the long-lifetime regime the lab frame decay probability scales as $P_{\rm dec} \propto \Umu m_N^6$, so the yield scales as $|U_{\mu N}|^4 m_N^6$.
A longer target raises the production rate, but also the accompanying rate of neutrino backgrounds and the distance HNLs must survive; this balance governs the reach of each configuration.
% Background-free projections follow the median expected CL$_s$ exclusion of the companion article~\cite{Read:2002hq}: $3.8$ signal events define the quoted $2\sigma$ contours.

The theoretical targets involve two quasi-degenerate HNLs, the minimum necessary to explain neutrino oscillation data and sufficient to seed leptogenesis in the early Universe~\cite{Asaka:2005pn,Hernandez:2015wna,Drewes:2016jae}.
To distinguish their mixings from the single-HNL benchmark we write $\Ual\equiv\sum_{I=2,3}|U_{\alpha N_I}|^2$ and $U^2=\sum_\alpha\Ual$. In the Casas--Ibarra parametrization~\cite{Casas:2001sr}, $U^2=(\sum_i m_i/m_N)\cosh(2\eta)$ up to corrections of order $\delta M/m_N$, with $\eta$ the imaginary part of the complex rotation angle.
The total Seesaw floor is found by taking $\eta = 0$ case, corresponding to $0.059~\mathrm{eV}/m_N$ for normal ordering (NO) and $0.099~\mathrm{eV}/m_N$ for inverted ordering (IO).
Simultaneously minimizing over $\eta$ and the Majorana phases gives the flavor-dependent Seesaw floor~\cite{Drewes:2016jae,Krasnov:2023jlt}
\begin{equation}\label{eq:flavor_floor}
    (\Ual)_{\rm min}
    =\frac{\left|m_a|V_{\alpha a}|^2-m_b|V_{\alpha b}|^2\right|}{m_N},
\end{equation}
where $V$ is the PMNS matrix and $(a,b)=(2,3)$ for NO or $(1,2)$ for IO. Interference between the two massive light states can drive a single flavor well below the total floor; at the NuFIT~5.3 best fit~\cite{Esteban:2020cvm} the IO muon-flavor floor lies below the NO one, reversing the ordering of the total floors.
%For simplicity, \cref{fig:combined} compares these pair-summed targets with the single-HNL sensitivities as benchmarks.
Both quasi-degenerate HNLs would be produced; establishing coverage of the seesaw targets requires combining their detectable decay yields with the correlated flavor mixings, lifetimes, and branching fractions, so the crossings shown in \cref{fig:combined} should be interpreted as benchmark comparisons.
%A model-specific exclusion would sum the yields of both HNLs with their correlated, flavor-dependent lifetimes and branching fractions.

\emph{Heavy neutrino factory.---}%
The first of our three proposals mirrors the neutrino factory concept~\cite{Albright:2000xi}.
A storage ring of muons with energy $E_\mu = 25~{\rm GeV}$ supplies a collimated HNL flux through muon decay for $2m_e<m_N<m_\mu$.
We consider a low-density, hodoscopic detector of length $100~\mathrm{m}$ and radius $5~\mathrm{m}$ aligned with a straight section, searching for $N\to\nu e^+e^-$, which has a branching ratio of about $10\%$.
The geometric acceptance is evaluated by sampling the rest-frame production spectrum and boosting to the laboratory; it grows with the detector radius and with $E_\mu$.
Assuming negligible background, the search improves on the LSND reinterpretation of Ref.~\cite{Ema:2023buz} by about two orders of magnitude.
This setup also has comparable sensitivity to \Ue~\cite{Arguelles:2026companion}.

Under standard cosmological assumptions, HNLs with these mixings are populated in the early Universe and, if their lifetime $\tau_N\propto(G_F^2U^2m_N^5)^{-1}$ is long enough for them to survive into big-bang nucleosynthesis (BBN), their decays, mainly into active neutrinos, raise the neutron-to-proton ratio and hence the primordial $^4$He abundance~\cite{Dolgov:2000jw,Sabti:2020yrt}.
This lifetime bound excludes the small-mixing and HNL mass region targeted by the heavy neutrino factory.
The search remains valuable in extensions beyond minimal mass-mixing that evade BBN bounds by shortening the HNL lifetime.
These include the addition of a transition magnetic moment, which enhances the decay width through the radiative channel $N \to \nu \gamma$ while production still proceeds through mixing~\cite{Arguelles:2021dqn}.
This scenario is discussed in the companion study~\cite{Arguelles:2026companion}.

\emph{Muon beam dump.---}%
The beam-dump proposal uses charged-current muon scattering in a target for HNL production and a low-density, hodoscopic downstream detector to observe HNL decays.
Neutrinos from muon decays in flight and from charged-current scattering in the target produce a beam-induced background: they interact in whatever medium fills the decay region, at a rate proportional to its density.
We consider three beam-dump configurations that cover different combinations of target and decay regions, offering tradeoffs between maximizing the HNL signal rate and minimizing neutrino backgrounds.
In all configurations we consider an incident muon energy of $5~{\rm TeV}$, though the $E_\mu = 1.5~\tev$ case is discussed in the companion article~\cite{Arguelles:2026companion}.

\emph{(a) Tungsten plus vacuum.}\
This configuration extends the proposal of Ref.~\cite{Cesarotti:2022ttv}.
The laboratory dump consists of a $40~\mathrm{m}$ tungsten target, a $10~\mathrm{m}$ magnetized shield assumed to suppress the residual beam and charged secondaries, and a $100~\mathrm{m}$ decay volume of radius $5~\mathrm{m}$.
A helium-filled volume at $1~\mathrm{atm}$, as in the NuTeV decay channel~\cite{NuTeV:1999kej,NuTeV:2001ndo}, would see $B\simeq2.5\times10^5$ neutrino interactions from a $5~\tev$ $\mu^-$ beam.
Evacuating the volume, following the SHiP design and NA62 operating strategy~\cite{SHiP:2021nfo,Miano:2021kyu,NA62:2023nhs,NA62:2023qyn}, suppresses neutrino interactions in the residual gas.
Achieving a neutrino background rate of $B<0.1$ requires
\begin{equation}\label{eq:vacP}
    P\lesssim4.1\times10^{-4}~\mathrm{mbar},
\end{equation}
between the SHiP specification ($\lesssim1~\mathrm{mbar}$) and the NA62 operating vacuum ($\sim10^{-6}~\mathrm{mbar}$). These experiments therefore provide precedents for vacuum decay volumes.
%the residual backgrounds of a multi-\tev\ muon dump require a dedicated assessment.

If a negligible neutrino background rate can be achieved, the sensitivity follows from a straightforward decay-counting experiment with no reconstruction efficiency factor applied.
The maximum $2\sigma$ reach of such an experiment is $\Umu = 2.8\times10^{-13}$ for $m_N \approx 40~\gev$.
% Its $2\sigma$ reach is $|U_{\mu N}|^2\simeq1.6\times10^{-12}$ near $m_N=22~\gev$ at $1.5~\tev$ and $2.8\times10^{-13}$ near $40~\gev$ at $5~\tev$.
In the companion article~\cite{Arguelles:2026companion} we consider several modifications to this configuration and assess their impact on the sensitivity.
A SHiP/NA62-style selection of exactly two charged tracks forming a vertex that points back to the target provides a reconstructible signature at some cost in rate: it reaches $8.0\times10^{-13}$ for $ m_N \approx 36~\gev$.
The $N\to\mu\pi$ channel dominates near $m_N\sim0.5~\gev$, while dilepton modes dominate at high masses.
A $10~\mathrm{m}$ water target~\cite{Arguelles:2026companion}  produces fewer HNLs and weakens the reach, but its smaller neutrino flux relaxes the pressure requirement to $2.6\times10^{-3}~\mathrm{mbar}$.
Finally, a helium-filled decay volume reduces the reach by about a factor of $20$ due to the high neutrino background rate.

\emph{(b) Earth plus atmosphere.}\ 
In this configuration, we direct the muon beam into bedrock to maximize the target mass for HNL production.
HNLs can then emerge from the Earth's surface and decay in the atmosphere, where the outgoing final state particles can be detected via Cherenkov light emission in air.
Our baseline steers the beam from $100~\mathrm{m}$ depth at an angle of $2.3^\circ$ above the horizon, giving a $2.4~\mathrm{km}$ path through the Earth.
Production is integrated over the degrading muon energy, using $-dE_\mu/dX=a+bE_\mu$ with $a/b\simeq571~\gev$~\cite{Groom:2001kq}, and the accumulated multiple Coulomb scattering is propagated to the HNL direction.

We search for $N\to\nu\mu^+\mu^-$, sampling the decay kinematics with \texttt{SIREN}~\cite{Schneider:2024eej}. Each muon must produce at least $12$ photoelectrons in an air-Cherenkov telescope that uses a Trinity-style SiPM camera of radius $4~\mathrm{m}$ and angular pixel size $0.3^\circ$~\cite{Eser:2023lck,POEMMA:2020ykm,Otte:2025dld,Otte:2018uxj}, and the two muons must be resolved in distinct pixels.
This signature removes the majority of neutrino backgrounds, though it can still be mimicked by $\nu_\mu$ charged-current charm production in the atmosphere with a muonic charm decay, where the remaining hadronic shower below the detection threshold.
We simulated this process in \texttt{SIREN} with cross sections drived from EPPS21NLO parton distributions functions on oxygen~\cite{Eskola:2021nhw}, Peterson fragmentation functions~\cite{Peterson:1982ak}, and PDG branching fractions of charm hadrons~\cite{ParticleDataGroup:2026mpi}.

\begin{figure}[t]
  \centering
  \includegraphics[width=\linewidth]{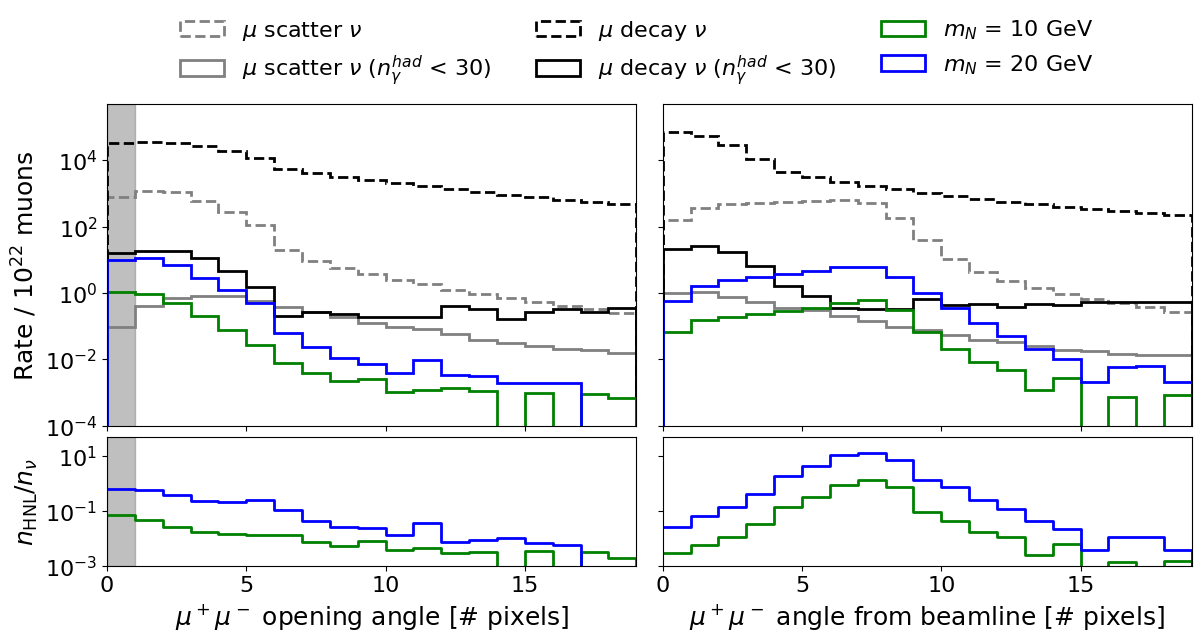}
  \caption{Dimuon opening angle (left) and angle to the beamline (right) in Dump~(b), in units of the $0.3^\circ$ camera pixels. Colored curves show HNL signals at $|U_{\mu N}|^2=10^{-12}$; black and gray curves show backgrounds from neutrinos produced in muon decay and scattering, before (dashed) and after (solid) the visible-shower veto. Lower panels show the signal-to-background ratio.}
  \label{fig:dimuon}
\end{figure}

\Cref{fig:dimuon} shows the dimuon opening angle $\Delta\phi_{\mu\mu}$ and the angle of the dimuon system to the beamline $\theta_{\mu\mu}$ for both HNLs and neutrino charm production, in units of $0.3^\circ$ camera pixels.
To separate signal and background, we require $0.3^\circ<\Delta\phi_{\mu\mu}<5^\circ$ and use a $\chi^2$ test statistic in $\theta_{\mu\mu}$ summed over the per-pixel bins in \cref{fig:dimuon}.
A longer atmospheric path increases the decay probability, but a shorter one improves the veto on the hadronic shower, so the best background-limited benchmark uses a $1~\mathrm{km}$ decay region with the camera at $50~\mathrm{m}$ altitude.
This setup reaches $|U_{\mu N}|^2\simeq9.9\times10^{-13}$ for $m_N \approx 26~\gev$.
An idealized background-free $20~\mathrm{km}$ region would reach the seesaw floor, though realizing it requires further investigation into detector geometries that can more efficiently reject neutrino-induced dimuons.

\emph{(c) Earth plus vacuum.}\ 
Pairing an Earth target of configuration (b) with the evacuated decay region of configuration (a) combines a kilometer-scale production volume with suppression of background neutrino interactions.
HNLs produced along a $1~\mathrm{km}$ beamline in rock are tracked to a $100~\mathrm{m}$ long, $5~\mathrm{m}$ radius evacuated tunnel.
The resulting sensitivity of a background-free decay-counting experiment reaches $|U_{\mu N}|^2\simeq1.8\times10^{-13}$ for $m_N\approx35~\gev$.
The larger neutrino flux from the rock requires
\begin{equation}\label{eq:vacP_tunnel}
    P\lesssim3.6\times10^{-6}~\mathrm{mbar}
\end{equation}
to keep the expected background below $0.1$ events, comparable to the NA62 vacuum~\cite{NA62:2023qyn}.
This requirement can be relaxed to approximately $10^{-4}~\mathrm{mbar}$ if the detector ignores the region within about $10~\mathrm{cm}$ of the beam axis, where decay-induced neutrinos concentrate, though this comes at the cost of some signal acceptance.

\emph{Muon--ion collider.---}%
Here we consider HNL production in muon-proton collisions~\cite{Acosta:2022ejc}.
A $7~\tev$ proton beam~\cite{Acosta:2022ejc,Akturk:2025ubm} provide $\sqrt{s}=2\sqrt{E_\mu E_p}\simeq6.5~(11.8)~\tev$ when paired with a muon beam of $E_\mu=1.5~(5)~\tev$.
For HNL detection, we require a displaced vertex between $0.4~\mathrm{mm}$ and $11~\mathrm{m}$, spanning the inner tracker to the outer muon system~\cite{Alimena:2019zri,Blondel:2022qqo}, and impose $m_N\gtrsim2~\gev$ as a trigger and reconstruction benchmark. With unit efficiency and negligible background the yield is $\sigma_0(m_N)\,|U_{\mu N}|^2\mathcal L\,P_{\rm dec}$, where $\sigma_0$ is the unit-mixing production cross section.

At $1.5~\tev$ with $1\abinv$ the reach is $|U_{\mu N}|^2\sim3\times10^{-9}$ near $m_N=15$--$20~\gev$; at $5~\tev$ with $10\abinv$ it improves to $2\times10^{-10}$ near $m_N = 30$--$35~\gev$. An optimistic $200\abinv$ benchmark reaches $\Umu \sim 1.4\times10^{-11}$ near $m_N = 56~\gev$, comparable to the FCC-ee displaced $\mu jj$ projection but weaker than its full-detector projection~\cite{Bellagamba:2025xpd}. The collider's strength is mass coverage: it retains sensitivity near $100~\gev$, beyond the on-shell $Z$-decay production used in those FCC-ee searches.

\emph{Combined reach and interpretation.---}%
Figure~\ref{fig:combined} summarizes the $2\sigma$ projections for the three proposals.
Their five benchmark contours provide continuous coverage for HNL masses from $2m_e$ to approximately $100~\gev$, with sensitivities well below existing limits and project limits from future experiments over most of this range, as shown in the left panel.
Below $m_\mu$ the heavy neutrino factory improves on existing limits by about two orders of magnitude; near $m_N\sim10~\gev$ the vacuum-based beam dump proposals lie up to six orders of magnitude below existing LHC limits, and about three below the projected HL-LHC sensitivity, and more than one below the projected FCC-ee sensitivity.
As shown in the right panel, Dump~(a) lies below the NO muon-flavor seesaw floor over $m_N\approx12$--$52~\gev$, by up to a factor of $2.5$, while Dump~(c) intersects the lower IO seesaw floor near $30~\gev$.
% These crossings compare idealized single-HNL decay yields with pair-summed theoretical targets.
% A model-specific exclusion requires the combined yields of both HNLs with their correlated flavor-dependent decay rates.
The other searches provide complementary capabilities: the heavy neutrino factory accesses MeV masses, the atmospheric configuration avoids an evacuated decay vessel, and the collider retains sensitivity near $100~\gev$.

The shaded bands in the right panel are the projected Akhmedov--Rubakov--Smirnov (ARS) leptogenesis targets~\cite{Akhmedov:1998qx}.
We solve momentum-averaged quantum kinetic equations~\cite{Drewes:2017zyw,Hernandez:2015wna,Ghiglieri:2017gjz} with \texttt{ULYSSES}~\cite{Granelli:2020pim,Granelli:2023knf}, scanning the mass splitting and CP phases at the NuFIT~5.3 best fit~\cite{Esteban:2020cvm}, with $\delta_{\rm CP}=197^\circ$ (NO) and $286^\circ$ (IO), and retaining points that generate $|\eta_B|\geq6.1\times10^{-10}$~\cite{Planck:2018vyg} before sphaleron freeze-out. Within the scan resolution, the lower edges of the bands approach the muon-flavor seesaw floors above $\sim1~\gev$.
The beam-dump proposals overlap these projected targets.
The IO muon-flavor target depends on the Dirac CP phase; the crossings therefore do not establish coverage for all allowed phases.

\emph{Conclusions.---}%
The three proposals use intense multi-\tev\ muon beams to search for the heavy particles that could generate neutrino mass, spanning HNL masses from the MeV scale to $100~\gev$.
The background-free muon beam dump proposals reach the NO and IO muon-flavor seesaw floors above $1~\gev$.
Their overlap with projected leptogenesis targets also connects these searches to mechanisms for generating the Universe's matter--antimatter asymmetry.

Realizing this reach requires validation of the vacuum detector, the charged-particle shielding, and near-unity visible-decay acceptance and reconstruction efficiency.
%In the long-lifetime regime an efficiency $\epsilon$ changes the reach in \Umu as $\epsilon^{-1/2}$.
Background uncertainties require a separate assessment; the atmospheric beam dump search in particular depends on the modeled hadronic Cherenkov emission and shower-veto efficiency.
In any case, coordination with the design and construction of the muon collider itself is essential.
Beam parameters including the muon energy, bunch frequency and dump rate have a strong on the sensitivities of these proposals.
HNL searches should therefore inform these choices from the outset.
If this happens, the muon collider could uncover the origin of neutrino masses and the matter-antimatter asymmetry of the Universe.

\begin{acknowledgments}
\emph{Acknowledgments.---}%
We thank the organizers of the ``Good $\nu$'s" workshop at Pitt PACC, during which initial discussions for this project took place.
We are grateful to Jackapan Pairin for providing a graphic design of our experimental concept.
The work of KX at MSU is supported in part by the U.S. National Science Foundation under Grants No. PHY-2112829, PHY-2013791, and PHY-2310497.
The work of KX at TDLI/SJTU is also supported by the National Natural Science Foundation of China and by the State Key Laboratory of Dark Matter Physics.
The work of P.-C.L.\ is supported by the National Natural Science Foundation of China under Grant No.~12405121.
The work of G.H. is supported by the Neutrino Theory Network Fellowship with contract number 726844. 
N.K. is supported by the National Science Foundation (NSF, CAREER Grant \#2239795) and the David \& Lucile Packard Foundation.
C.A.A. is supported by the Faculty of Arts and Sciences of Harvard University, the National Science Foundation (NSF, CAREER Grant \#2239795), the John Templeton Foundation (Grant \#63651), the Research Corporation for Science Advancement, and the David \& Lucile Packard Foundation.
N.K. and C.A.A. are supported by the National Science Foundation under Cooperative Agreement PHY-2019786 (The NSF AI Institute for Artificial Intelligence and Fundamental Interactions, http://iaifi.org/).
This work made use of \texttt{MadGraph5\_aMC@NLO}~\cite{Alwall:2014hca}, the CT18NNLO parton distributions~\cite{Hou:2019efy}, \texttt{SIREN}~\cite{Schneider:2024eej}, and \texttt{ULYSSES}~\cite{Granelli:2020pim,Granelli:2023knf}.
\end{acknowledgments}

\bibliography{bib}

@article{Denton:2024glz,
    author = "Denton, Peter B. and Gehrlein, Julia",
    title = "{A modern look at the oscillation physics case for a neutrino factory}",
    eprint = "2407.02572",
    archivePrefix = "arXiv",
    primaryClass = "hep-ph",
    doi = "10.1016/j.nuclphysb.2025.116818",
    journal = "Nucl. Phys. B",
    volume = "1012",
    pages = "116818",
    year = "2025"
}

@article{Hou:2019efy,
    author = "Hou, Tie-Jiun and others",
    title = "{New CTEQ global analysis of quantum chromodynamics with high-precision data from the LHC}",
    eprint = "1912.10053",
    archivePrefix = "arXiv",
    primaryClass = "hep-ph",
    reportNumber = "MSUHEP-19-025, PITT-PACC-1911, SMU-HEP-19-03",
    doi = "10.1103/PhysRevD.103.014013",
    journal = "Phys. Rev. D",
    volume = "103",
    number = "1",
    pages = "014013",
    year = "2021"
}

@article{Alwall:2014hca,
    author = "Alwall, J. and Frederix, R. and Frixione, S. and Hirschi, V. and Maltoni, F. and Mattelaer, O. and Shao, H. -S. and Stelzer, T. and Torrielli, P. and Zaro, M.",
    title = "{The automated computation of tree-level and next-to-leading order differential cross sections, and their matching to parton shower simulations}",
    eprint = "1405.0301",
    archivePrefix = "arXiv",
    primaryClass = "hep-ph",
    reportNumber = "CERN-PH-TH-2014-064, CP3-14-18, LPN14-066, MCNET-14-09, ZU-TH-14-14",
    doi = "10.1007/JHEP07(2014)079",
    journal = "JHEP",
    volume = "07",
    pages = "079",
    year = "2014"
}

@article{Ibarra:2011xn,
    author = "Ibarra, A. and Molinaro, E. and Petcov, S. T.",
    title = "{Low Energy Signatures of the TeV Scale See-Saw Mechanism}",
    eprint = "1103.6217",
    archivePrefix = "arXiv",
    primaryClass = "hep-ph",
    doi = "10.1103/PhysRevD.84.013005",
    journal = "Phys. Rev. D",
    volume = "84",
    pages = "013005",
    year = "2011"
}

@article{Atre:2009rg,
    author = "Atre, Anupama and Han, Tao and Pascoli, Silvia and Zhang, Bin",
    title = "{The Search for Heavy Majorana Neutrinos}",
    eprint = "0901.3589",
    archivePrefix = "arXiv",
    primaryClass = "hep-ph",
    reportNumber = "FERMILAB-PUB-08-086-T, NSF-KITP-08-54, MADPH-06-1466, DCPT-07-198, IPPP-07-99",
    doi = "10.1088/1126-6708/2009/05/030",
    journal = "JHEP",
    volume = "05",
    pages = "030",
    year = "2009"
}

@article{Drewes:2016jae,
    author = "Drewes, Marco and Garbrecht, Bjorn and Gueter, Dario and Klaric, Juraj",
    title = "{Testing the low scale seesaw and leptogenesis}",
    eprint = "1609.09069",
    archivePrefix = "arXiv",
    primaryClass = "hep-ph",
    reportNumber = "TUM-HEP-1062-16",
    doi = "10.1007/JHEP08(2017)018",
    journal = "JHEP",
    volume = "08",
    pages = "018",
    year = "2017"
}

@article{Accettura:2023ked,
    author = "Accettura, Carlotta and others",
    title = "{Towards a muon collider}",
    eprint = "2303.08533",
    archivePrefix = "arXiv",
    primaryClass = "physics.acc-ph",
    reportNumber = "FERMILAB-PUB-23-123-AD-PPD-T",
    doi = "10.1140/epjc/s10052-023-11889-x",
    journal = "Eur. Phys. J. C",
    volume = "83",
    number = "9",
    pages = "864",
    year = "2023",
    note = "[Erratum: Eur.Phys.J.C 84, 36 (2024)]"
}

@misc{InternationalMuonCollider:2025sys,
    author = "Accettura, Carlotta and others",
    collaboration = "International Muon Collider",
    title = "{The Muon Collider}",
    eprint = "2504.21417",
    archivePrefix = "arXiv",
    primaryClass = "physics.acc-ph",
    reportNumber = "FERMILAB-PUB-25-0309-AD-PPD-T",
    month = "4",
    year = "2025"
}

@article{AlAli:2021let,
    author = "Al Ali, Hind and others",
    title = "{The muon Smasher{\textquoteright}s guide}",
    eprint = "2103.14043",
    archivePrefix = "arXiv",
    primaryClass = "hep-ph",
    doi = "10.1088/1361-6633/ac6678",
    journal = "Rept. Prog. Phys.",
    volume = "85",
    number = "8",
    pages = "084201",
    year = "2022"
}

@article{Geer:1997iz,
    author = "Geer, S.",
    title = "{Neutrino beams from muon storage rings: Characteristics and physics potential}",
    eprint = "hep-ph/9712290",
    archivePrefix = "arXiv",
    reportNumber = "FERMILAB-PUB-97-389",
    doi = "10.1103/PhysRevD.57.6989",
    journal = "Phys. Rev. D",
    volume = "57",
    pages = "6989--6997",
    year = "1998",
    note = "[Erratum: Phys.Rev.D 59, 039903 (1999)]"
}

@article{Bojorquez-Lopez:2024bsr,
    author = {Bojorquez-Lopez, Luc and Hostert, Matheus and Arg{\"u}elles, Carlos A. and Liu, Zhen},
    title = "{Muon Colliders and the Neutrino Slice}",
    eprint = "2412.14115",
    archivePrefix = "arXiv",
    primaryClass = "hep-ph",
    reportNumber = "UMN-TH-4409/24",
    doi = "10.1103/4sj1-ycft",
    journal = "Phys. Rev. Lett.",
    volume = "135",
    number = "9",
    pages = "091803",
    year = "2025"
}

@misc{Mangano:2001mj,
    author = "Mangano, Michelangelo L. and others",
    editor = "Blondel, A. and others",
    title = "{Physics at the front end of a neutrino factory: A Quantitative appraisal}",
    eprint = "hep-ph/0105155",
    archivePrefix = "arXiv",
    reportNumber = "CERN-TH-2001-131",
    doi = "10.5170/CERN-2004-002.185",
    pages = "185--257",
    month = "5",
    year = "2001"
}

@misc{Akturk:2025ubm,
    author = "Akturk, D. and others",
    title = "{$\mu$LHC: Antimuon Ring and HL-LHC based $\mu^+p$ Collider}",
    eprint = "2506.15445",
    archivePrefix = "arXiv",
    primaryClass = "physics.acc-ph",
    month = "6",
    year = "2025"
}

@article{Acosta:2022ejc,
    author = "Acosta, Darin and Barberis, Emanuela and Hurley, Nicholas and Li, Wei and Miguel Colin, Osvaldo and Wang, Yijie and Wood, Darien and Zuo, Xunwu",
    title = "{The potential of a TeV-scale muon-ion collider}",
    eprint = "2203.06258",
    archivePrefix = "arXiv",
    primaryClass = "hep-ex",
    doi = "10.1088/1748-0221/18/09/P09025",
    journal = "JINST",
    volume = "18",
    number = "09",
    pages = "P09025",
    year = "2023"
}

@article{Alimena:2019zri,
    author = "Alimena, Juliette and others",
    title = "{Searching for long-lived particles beyond the Standard Model at the Large Hadron Collider}",
    eprint = "1903.04497",
    archivePrefix = "arXiv",
    primaryClass = "hep-ex",
    doi = "10.1088/1361-6471/ab4574",
    journal = "J. Phys. G",
    volume = "47",
    number = "9",
    pages = "090501",
    year = "2020"
}

@article{Akhmedov:1998qx,
    author = "Akhmedov, E. Kh. and Rubakov, V. A. and Smirnov, A. Yu.",
    title = "{Baryogenesis via neutrino oscillations}",
    eprint = "hep-ph/9803255",
    archivePrefix = "arXiv",
    reportNumber = "IC-98-22",
    doi = "10.1103/PhysRevLett.81.1359",
    journal = "Phys. Rev. Lett.",
    volume = "81",
    pages = "1359--1362",
    year = "1998"
}

@article{Asaka:2005pn,
    author = "Asaka, Takehiko and Shaposhnikov, Mikhail",
    title = "{The $\nu$MSM, dark matter and baryon asymmetry of the universe}",
    eprint = "hep-ph/0505013",
    archivePrefix = "arXiv",
    doi = "10.1016/j.physletb.2005.06.020",
    journal = "Phys. Lett. B",
    volume = "620",
    pages = "17--26",
    year = "2005"
}

@article{Casas:2001sr,
    author = "Casas, J. A. and Ibarra, A.",
    title = "{Oscillating neutrinos and $\mu \to e, \gamma$}",
    eprint = "hep-ph/0103065",
    archivePrefix = "arXiv",
    reportNumber = "IEM-FT-211-01, OUTP-01-11P, IFT-UAM-CSIC-01-08",
    doi = "10.1016/S0550-3213(01)00475-8",
    journal = "Nucl. Phys. B",
    volume = "618",
    pages = "171--204",
    year = "2001"
}

@article{Esteban:2020cvm,
    author = "Esteban, Ivan and Gonzalez-Garcia, M. C. and Maltoni, Michele and Schwetz, Thomas and Zhou, Albert",
    title = "{The fate of hints: updated global analysis of three-flavor neutrino oscillations}",
    eprint = "2007.14792",
    archivePrefix = "arXiv",
    primaryClass = "hep-ph",
    doi = "10.1007/JHEP09(2020)178",
    journal = "JHEP",
    volume = "09",
    pages = "178",
    year = "2020"
}

@article{Granelli:2020pim,
    author = "Granelli, Alessandro and Moffat, Kristian and Perez-Gonzalez, Yuber F. and Schulz, Holger and Turner, Jessica",
    title = "{ULYSSES: Universal LeptogeneSiS Equation Solver}",
    eprint = "2007.09150",
    archivePrefix = "arXiv",
    primaryClass = "hep-ph",
    reportNumber = "FERMILAB-PUB-20-275-T, SISSA 17/2020/FISI, IPPP/20/30",
    doi = "10.1016/j.cpc.2020.107813",
    journal = "Comput. Phys. Commun.",
    volume = "262",
    pages = "107813",
    year = "2021"
}

@article{Granelli:2023knf,
    author = "Granelli, Alessandro and Leslie, Christopher and Perez-Gonzalez, Yuber F. and Schulz, Holger and Shuve, Brian and Turner, Jessica and Walker, Rosie",
    title = "{ULYSSES, Universal LeptogeneSiS Equation Solver: version 2}",
    eprint = "2301.05722",
    archivePrefix = "arXiv",
    primaryClass = "hep-ph",
    doi = "10.1016/j.cpc.2023.108834",
    journal = "Comput. Phys. Commun.",
    volume = "291",
    pages = "108834",
    year = "2023"
}

@article{Ghiglieri:2017gjz,
    author = "Ghiglieri, J. and Laine, M.",
    title = "{GeV-scale hot sterile neutrino oscillations: a numerical solution}",
    eprint = "1711.08469",
    archivePrefix = "arXiv",
    primaryClass = "hep-ph",
    doi = "10.1007/JHEP02(2018)078",
    journal = "JHEP",
    volume = "02",
    pages = "078",
    year = "2018"
}

@article{Planck:2018vyg,
    author = "Aghanim, N. and others",
    collaboration = "Planck",
    title = "{Planck 2018 results. VI. Cosmological parameters}",
    eprint = "1807.06209",
    archivePrefix = "arXiv",
    primaryClass = "astro-ph.CO",
    doi = "10.1051/0004-6361/201833910",
    journal = "Astron. Astrophys.",
    volume = "641",
    pages = "A6",
    year = "2020",
    note = "[Erratum: Astron.Astrophys. 652, C4 (2021)]"
}

@article{Cesarotti:2022ttv,
    author = "Cesarotti, Cari and Homiller, Samuel and Mishra, Rashmish K. and Reece, Matthew",
    title = "{Probing New Gauge Forces with a High-Energy Muon Beam Dump}",
    eprint = "2202.12302",
    archivePrefix = "arXiv",
    primaryClass = "hep-ph",
    doi = "10.1103/PhysRevLett.130.071803",
    journal = "Phys. Rev. Lett.",
    volume = "130",
    number = "7",
    pages = "071803",
    year = "2023"
}

@article{Drewes:2017zyw,
    author = "Drewes, M. and Garbrecht, B. and Hernandez, P. and Kekic, M. and Lopez-Pavon, J. and Racker, J. and Rius, N. and Salvado, J. and Teresi, D.",
    title = "{ARS Leptogenesis}",
    eprint = "1711.02862",
    archivePrefix = "arXiv",
    primaryClass = "hep-ph",
    doi = "10.1142/S0217751X18420022",
    journal = "Int. J. Mod. Phys. A",
    volume = "33",
    number = "05n06",
    pages = "1842002",
    year = "2018"
}

@article{Hernandez:2015wna,
    author = "Hern{\'a}ndez, P. and Kekic, M. and L{\'o}pez-Pav{\'o}n, J. and Racker, J. and Rius, N.",
    title = "{Leptogenesis in GeV scale seesaw models}",
    eprint = "1508.03676",
    archivePrefix = "arXiv",
    primaryClass = "hep-ph",
    reportNumber = "IFIC-15-44, SISSA-35-2015-FISI",
    doi = "10.1007/JHEP10(2015)067",
    journal = "JHEP",
    volume = "10",
    pages = "067",
    year = "2015"
}

@article{Abdullahi:2022jlv,
    author = "Abdullahi, Asli M. and others",
    title = "{The present and future status of heavy neutral leptons}",
    eprint = "2203.08039",
    archivePrefix = "arXiv",
    primaryClass = "hep-ph",
    reportNumber = "FERMILAB-CONF-22-184-T-V",
    doi = "10.1088/1361-6471/ac98f9",
    journal = "J. Phys. G",
    volume = "50",
    number = "2",
    pages = "020501",
    year = "2023"
}

@article{Ema:2023buz,
    author = "Ema, Yohei and Liu, Zhen and Lyu, Kun-Feng and Pospelov, Maxim",
    title = "{Heavy Neutral Leptons from Stopped Muons and Pions}",
    eprint = "2306.07315",
    archivePrefix = "arXiv",
    primaryClass = "hep-ph",
    reportNumber = "UMN-TH-4217/23, FTPI-MINN-23-10",
    doi = "10.1007/JHEP08(2023)169",
    journal = "JHEP",
    volume = "08",
    pages = "169",
    year = "2023"
}

@article{Breitbach:2021gvv,
    author = "Breitbach, Moritz and Buonocore, Luca and Frugiuele, Claudia and Kopp, Joachim and Mittnacht, Lukas",
    title = "{Searching for physics beyond the Standard Model in an off-axis DUNE near detector}",
    eprint = "2102.03383",
    archivePrefix = "arXiv",
    primaryClass = "hep-ph",
    reportNumber = "MITP-21-004, ZU-TH-5/21, CERN-TH-2021-018",
    doi = "10.1007/JHEP01(2022)048",
    journal = "JHEP",
    volume = "01",
    pages = "048",
    year = "2022"
}

@article{SHiP:2018xqw,
    author = "Ahdida, C. and others",
    collaboration = "SHiP",
    title = "{Sensitivity of the SHiP experiment to Heavy Neutral Leptons}",
    eprint = "1811.00930",
    archivePrefix = "arXiv",
    primaryClass = "hep-ph",
    doi = "10.1007/JHEP04(2019)077",
    journal = "JHEP",
    volume = "04",
    pages = "077",
    year = "2019"
}

@misc{PIONEER:2022yag,
    author = "Altmannshofer, W. and others",
    collaboration = "PIONEER",
    title = "{PIONEER: Studies of Rare Pion Decays}",
    eprint = "2203.01981",
    archivePrefix = "arXiv",
    primaryClass = "hep-ex",
    month = "3",
    year = "2022"
}

@article{Drewes:2019fou,
    author = "Drewes, Marco and Hajer, Jan",
    title = "{Heavy Neutrinos in displaced vertex searches at the LHC and HL-LHC}",
    eprint = "1903.06100",
    archivePrefix = "arXiv",
    primaryClass = "hep-ph",
    reportNumber = "CP3-19-11",
    doi = "10.1007/JHEP02(2020)070",
    journal = "JHEP",
    volume = "02",
    pages = "070",
    year = "2020"
}

@article{Blondel:2022qqo,
    author = "Blondel, A. and others",
    title = "{Searches for long-lived particles at the future FCC-ee}",
    eprint = "2203.05502",
    archivePrefix = "arXiv",
    primaryClass = "hep-ex",
    doi = "10.3389/fphy.2022.967881",
    journal = "Front. in Phys.",
    volume = "10",
    pages = "967881",
    year = "2022"
}

@article{Bellagamba:2025xpd,
    author = "Bellagamba, L. and Polesello, G. and Valle, N.",
    title = "{Searches for heavy neutral leptons at FCC-ee in final states including a muon}",
    eprint = "2503.19464",
    archivePrefix = "arXiv",
    primaryClass = "hep-ex",
    doi = "10.1140/epjc/s10052-025-14749-y",
    journal = "Eur. Phys. J. C",
    volume = "85",
    number = "9",
    pages = "1069",
    year = "2025"
}

@article{NuTeV:1999kej,
    author = "Vaitaitis, A. and others",
    collaboration = "NuTeV, E815",
    title = "{Search for neutral heavy leptons in a high-energy neutrino beam}",
    eprint = "hep-ex/9908011",
    archivePrefix = "arXiv",
    reportNumber = "FERMILAB-PUB-99-223-E",
    doi = "10.1103/PhysRevLett.83.4943",
    journal = "Phys. Rev. Lett.",
    volume = "83",
    pages = "4943--4946",
    year = "1999"
}

@article{Arguelles:2021dqn,
    author = {Arg{\"u}elles, Carlos A. and Foppiani, Nicol{\`o} and Hostert, Matheus},
    title = "{Heavy neutral leptons below the kaon mass at hodoscopic neutrino detectors}",
    eprint = "2109.03831",
    archivePrefix = "arXiv",
    primaryClass = "hep-ph",
    doi = "10.1103/PhysRevD.105.095006",
    journal = "Phys. Rev. D",
    volume = "105",
    number = "9",
    pages = "095006",
    year = "2022"
}

@article{Minkowski:1977sc,
    author = "Minkowski, Peter",
    title = "{$\mu \to e\gamma$ at a Rate of One Out of $10^{9}$ Muon Decays?}",
    reportNumber = "Print-77-0182 (BERN)",
    doi = "10.1016/0370-2693(77)90435-X",
    journal = "Phys. Lett. B",
    volume = "67",
    pages = "421--428",
    year = "1977"
}

@article{Mohapatra:1979ia,
    author = "Mohapatra, Rabindra N. and Senjanovic, Goran",
    title = "{Neutrino Mass and Spontaneous Parity Nonconservation}",
    reportNumber = "MDDP-TR-80-060, MDDP-PP-80-105, CCNY-HEP-79-10",
    doi = "10.1103/PhysRevLett.44.912",
    journal = "Phys. Rev. Lett.",
    volume = "44",
    pages = "912",
    year = "1980"
}

@article{Gell-Mann:1979vob,
    author = "Gell-Mann, Murray and Ramond, Pierre and Slansky, Richard",
    title = "{Complex Spinors and Unified Theories}",
    eprint = "1306.4669",
    archivePrefix = "arXiv",
    primaryClass = "hep-th",
    reportNumber = "PRINT-80-0576",
    journal = "Conf. Proc. C",
    volume = "790927",
    pages = "315--321",
    year = "1979"
}

@article{Yanagida:1979as,
    author = "Yanagida, Tsutomu",
    editor = "Sawada, Osamu and Sugamoto, Akio",
    title = "{Horizontal gauge symmetry and masses of neutrinos}",
    reportNumber = "KEK-79-18-95",
    journal = "Conf. Proc. C",
    volume = "7902131",
    pages = "95--99",
    year = "1979"
}

@article{Kuzmin:1985mm,
    author = "Kuzmin, V. A. and Rubakov, V. A. and Shaposhnikov, M. E.",
    title = "{On the Anomalous Electroweak Baryon Number Nonconservation in the Early Universe}",
    reportNumber = "IC/85/8",
    doi = "10.1016/0370-2693(85)91028-7",
    journal = "Phys. Lett. B",
    volume = "155",
    pages = "36",
    year = "1985"
}

@article{POEMMA:2020ykm,
    author = "Olinto, A. V. and others",
    collaboration = "POEMMA",
    title = "{The POEMMA (Probe of Extreme Multi-Messenger Astrophysics) observatory}",
    eprint = "2012.07945",
    archivePrefix = "arXiv",
    primaryClass = "astro-ph.IM",
    doi = "10.1088/1475-7516/2021/06/007",
    journal = "JCAP",
    volume = "06",
    pages = "007",
    year = "2021"
}

@article{Eser:2023lck,
    author = "Eser, Johannes and Olinto, Angela V. and Wiencke, Lawrence",
    collaboration = "JEM-EUSO",
    title = "{Overview and First Results of EUSO-SPB2}",
    eprint = "2308.15693",
    archivePrefix = "arXiv",
    primaryClass = "astro-ph.HE",
    doi = "10.22323/1.444.0397",
    journal = "PoS",
    volume = "ICRC2023",
    pages = "397",
    year = "2023"
}

@article{Otte:2025dld,
    author = "Otte, A. Nepomuk",
    collaboration = "Trinity",
    title = "{The Trinity-One PeV-Neutrino Telescope}",
    eprint = "2509.18237",
    archivePrefix = "arXiv",
    primaryClass = "astro-ph.HE",
    doi = "10.22323/1.501.1136",
    journal = "PoS",
    volume = "ICRC2025",
    pages = "1136",
    year = "2025"
}

@article{Groom:2001kq,
    author = "Groom, Donald E. and Mokhov, Nikolai V. and Striganov, Sergei I.",
    title = "{Muon stopping power and range tables 10-MeV to 100-TeV}",
    doi = "10.1006/adnd.2001.0861",
    journal = "Atom. Data Nucl. Data Tabl.",
    volume = "78",
    pages = "183--356",
    year = "2001"
}

@misc{Burk:2026fox,
    author = "Burk, Francis M. and Han, Tao and Kilian, Wolfgang and Kling, Felix and Kopp, Joachim and Tabrizi, Zahra",
    title = "{Neutrino Fluxes at a Muon Collider}",
    eprint = "2608.02714",
    archivePrefix = "arXiv",
    primaryClass = "hep-ph",
    month = "8",
    year = "2026"
}

@article{InternationalMuonCollider:2024jyv,
    author = "Accettura, Carlotta and others",
    collaboration = "International Muon Collider",
    title = "{Interim report for the International Muon Collider Collaboration (IMCC)}",
    eprint = "2407.12450",
    archivePrefix = "arXiv",
    primaryClass = "physics.acc-ph",
    reportNumber = "CERN-2024-002",
    doi = "10.23731/CYRM-2024-002",
    journal = "CERN Yellow Rep. Monogr.",
    volume = "2/2024",
    pages = "176",
    year = "2024"
}

@article{SHiP:2021nfo,
    author = "Ahdida, C. and others",
    collaboration = "SHiP",
    title = "{The SHiP experiment at the proposed CERN SPS Beam Dump Facility}",
    eprint = "2112.01487",
    archivePrefix = "arXiv",
    primaryClass = "physics.ins-det",
    doi = "10.1140/epjc/s10052-022-10346-5",
    journal = "Eur. Phys. J. C",
    volume = "82",
    number = "5",
    pages = "486",
    year = "2022"
}

@article{Miano:2021kyu,
    author = "Miano, Andrea and Fiorillo, Andrea and Salzano, Antonio and Prota, Andrea and Jacobsson, Richard",
    title = "{The structural design of the decay volume for the Search for Hidden Particles (SHIP) project}",
    doi = "10.1007/s43452-020-00152-9",
    journal = "Archiv.Civ.Mech.Eng.",
    volume = "21",
    number = "1",
    pages = "3",
    year = "2021"
}

@article{NA62:2023nhs,
    author = "Cortina Gil, Eduardo and others",
    collaboration = "NA62",
    title = "{Search for Leptonic Decays of Dark Photons at NA62}",
    eprint = "2312.12055",
    archivePrefix = "arXiv",
    primaryClass = "hep-ex",
    reportNumber = "CERN-EP-2023-296",
    doi = "10.1103/PhysRevLett.133.111802",
    journal = "Phys. Rev. Lett.",
    volume = "133",
    number = "11",
    pages = "111802",
    year = "2024"
}

@article{NA62:2023qyn,
    author = "Cortina Gil, Eduardo and others",
    collaboration = "NA62",
    title = "{Search for dark photon decays to $\mu^{+}\mu^{-}$ at NA62}",
    eprint = "2303.08666",
    archivePrefix = "arXiv",
    primaryClass = "hep-ex",
    reportNumber = "CERN-EP-2023-032",
    doi = "10.1007/JHEP09(2023)035",
    journal = "JHEP",
    volume = "09",
    pages = "035",
    year = "2023"
}

@article{NuTeV:2001ndo,
    author = "Adams, T. and others",
    collaboration = "NuTeV",
    title = "{Observation of an Anomalous Number of Dimuon Events in a High-Energy Neutrino Beam}",
    eprint = "hep-ex/0104037",
    archivePrefix = "arXiv",
    reportNumber = "FERMILAB-PUB-01-0560-PPD",
    doi = "10.1103/PhysRevLett.87.041801",
    journal = "Phys. Rev. Lett.",
    volume = "87",
    pages = "041801",
    year = "2001"
}

@article{Arguelles:2026companion,
    author = {Arg{\"u}elles, Carlos A. and Herrera, Gonzalo and Kamp, Nicholas and Lu, Peng-Cheng and Xie, Keping},
    title = "{Novel Heavy Neutral Lepton Searches With Next-Generation Muon Accelerators}",
    journal = "to appear",
    year = "2026",
    note = "companion paper"
}

@article{Sabti:2020yrt,
    author = "Sabti, Nashwan and Magalich, Andrii and Filimonova, Anastasiia",
    title = "{An Extended Analysis of Heavy Neutral Leptons during Big Bang Nucleosynthesis}",
    eprint = "2006.07387",
    archivePrefix = "arXiv",
    primaryClass = "hep-ph",
    reportNumber = "KCL-2020-09",
    doi = "10.1088/1475-7516/2020/11/056",
    journal = "JCAP",
    volume = "11",
    pages = "056",
    year = "2020"
}

@article{Fernandez-Martinez:2023phj,
    author = "Fern\'andez-Mart\'\i{}nez, Enrique and Gonz\'alez-L\'opez, Manuel and Hern\'andez-Garc\'\i{}a, Josu and Hostert, Matheus and L\'opez-Pav\'on, Jacobo",
    title = "{Effective portals to heavy neutral leptons}",
    eprint = "2304.06772",
    archivePrefix = "arXiv",
    primaryClass = "hep-ph",
    reportNumber = "FTUV-23-0303.1224, IFIC/23-09",
    doi = "10.1007/JHEP09(2023)001",
    journal = "JHEP",
    volume = "09",
    pages = "001",
    year = "2023"
}

@misc{Drewes:2019mhg,
    author = "Drewes, Marco",
    title = "{On the Minimal Mixing of Heavy Neutrinos}",
    eprint = "1904.11959",
    archivePrefix = "arXiv",
    primaryClass = "hep-ph",
    reportNumber = "CP3-19-20",
    month = "4",
    year = "2019"
}

@misc{Krasnov:2023jlt,
    author = "Krasnov, Igor",
    title = "{HNL see-saw: lower mixing limit and pseudodegenerate state}",
    eprint = "2307.01190",
    archivePrefix = "arXiv",
    primaryClass = "hep-ph",
    month = "7",
    year = "2023"
}

@article{Eskola:2021nhw,
    author = "Eskola, Kari J. and Paakkinen, Petja and Paukkunen, Hannu and Salgado, Carlos A.",
    title = "{EPPS21: a global QCD analysis of nuclear PDFs}",
    eprint = "2112.12462",
    archivePrefix = "arXiv",
    primaryClass = "hep-ph",
    doi = "10.1140/epjc/s10052-022-10359-0",
    journal = "Eur. Phys. J. C",
    volume = "82",
    number = "5",
    pages = "413",
    year = "2022"
}

@article{Schneider:2024eej,
    author = "Schneider, Austin and Kamp, Nicholas W. and Wen, Alex Y.",
    title = "{SIREN: An open-source neutrino injection toolkit}",
    eprint = "2406.01745",
    archivePrefix = "arXiv",
    primaryClass = "hep-ex",
    doi = "10.1016/j.cpc.2025.109799",
    journal = "Comput. Phys. Commun.",
    volume = "316",
    pages = "109799",
    year = "2025"
}

@article{Peterson:1982ak,
    author = "Peterson, C. and Schlatter, D. and Schmitt, I. and Zerwas, Peter M.",
    title = "{Scaling Violations in Inclusive e+ e- Annihilation Spectra}",
    reportNumber = "SLAC-PUB-2912",
    doi = "10.1103/PhysRevD.27.105",
    journal = "Phys. Rev. D",
    volume = "27",
    pages = "105",
    year = "1983"
}

@article{ParticleDataGroup:2026mpi,
    author = "Takahashi, F. and others",
    collaboration = "Particle Data Group",
    title = "{Review of Particle Physics}",
    doi = "10.1142/s0217751x26300115",
    journal = "Int. J. Mod. Phys. A",
    volume = "41",
    number = "22",
    pages = "2630011",
    year = "2026"
}

@article{Otte:2018uxj,
    author = "Otte, Adam Nepomuk",
    title = "{Studies of an air-shower imaging system for the detection of ultrahigh-energy neutrinos}",
    eprint = "1811.09287",
    archivePrefix = "arXiv",
    primaryClass = "astro-ph.IM",
    doi = "10.1103/PhysRevD.99.083012",
    journal = "Phys. Rev. D",
    volume = "99",
    number = "8",
    pages = "083012",
    year = "2019"
}

@misc{Baldini:2021shadows,
    author = "Baldini, W. and others",
    title = "{SHADOWS (Search for Hidden And Dark Objects With the SPS)}",
    eprint = "2110.08025",
    archivePrefix = "arXiv",
    primaryClass = "hep-ex",
    month = "10",
    year = "2021"
}

@misc{Choi:2026yzw,
    author = "Choi, Ju-Yeol and Hostert, Matheus and Li, Peiran and Liu, Zhen",
    title = "{The Forward Neutrino Flux and its Secondaries at a 10 TeV Muon Collider}",
    eprint = "2608.02718",
    archivePrefix = "arXiv",
    primaryClass = "hep-ph",
    month = "8",
    year = "2026"
}

@misc{Albright:2000xi,
    author = "Albright, C. and others",
    title = "{Physics at a Neutrino Factory}",
    eprint = "hep-ex/0008064",
    archivePrefix = "arXiv",
    reportNumber = "FERMILAB-FN-0692",
    month = "8",
    year = "2000"
}

@article{Fukugita:1986hr,
    author = "Fukugita, M. and Yanagida, T.",
    title = "{Baryogenesis Without Grand Unification}",
    reportNumber = "RIFP-641",
    doi = "10.1016/0370-2693(86)91126-3",
    journal = "Phys. Lett. B",
    volume = "174",
    pages = "45--47",
    year = "1986"
}

@article{Gorbunov:2007ak,
    author = "Gorbunov, Dmitry and Shaposhnikov, Mikhail",
    title = "{How to find neutral leptons of the $\nu$MSM?}",
    eprint = "0705.1729",
    archivePrefix = "arXiv",
    primaryClass = "hep-ph",
    doi = "10.1088/1126-6708/2007/10/015",
    journal = "JHEP",
    volume = "10",
    pages = "015",
    year = "2007",
    note = "[Erratum: JHEP 11, 101 (2013)]"
}

@misc{Begel:2025ldu,
    author = "Begel, M. and others",
    title = "{United States Muon Collider Community White Paper for the European Strategy for Particle Physics Update}",
    eprint = "2503.23695",
    archivePrefix = "arXiv",
    primaryClass = "hep-ex",
    reportNumber = "FERMILAB-PUB-25-0203-AD",
    month = "3",
    year = "2025"
}

@article{Dolgov:2000jw,
    author = "Dolgov, A. D. and Hansen, S. H. and Raffelt, G. and Semikoz, D. V.",
    title = "{Heavy sterile neutrinos: Bounds from big bang nucleosynthesis and SN1987A}",
    eprint = "hep-ph/0008138",
    archivePrefix = "arXiv",
    doi = "10.1016/S0550-3213(00)00566-6",
    journal = "Nucl. Phys. B",
    volume = "590",
    pages = "562--574",
    year = "2000"
}

\end{document}